\documentclass[aps,prl,twocolumn,showpacs,groupedaddress]{revtex4}

\usepackage{graphicx}% Include figure files
\usepackage{dcolumn}% Align table columns on decimal point
\usepackage{bm}% bold math
\usepackage{hyperref}% add hypertext capabilities
\usepackage{amsmath}
\usepackage[usenames]{color}
\begin{document}

\title{Centrality scaling in large networks}

\author{M\'aria Ercsey-Ravasz}
 \email{mercseyr@nd.edu}
\author{Zolt\'an Toroczkai}
 \email{toro@nd.edu}
\affiliation{Interdisciplinary Center for Network Science and Applications (iCeNSA), 
Department of Physics, \\ University of Notre Dame, Notre Dame, IN, 46556 USA
}

\date{\today}

\begin{abstract}
Betweenness centrality lies at the core of both transport and structural vulnerability 
properties of complex networks, however, it is computationally costly, and its 
measurement for networks with millions of nodes is  nearly impossible. By introducing 
a multiscale decomposition of shortest paths, we show that the contributions to 
betweenness coming from geodesics not longer than $L$ obey a characteristic scaling 
vs $L$, which can be used to predict the distribution of the full centralities. 
The method is also illustrated on a real-world social network of $5.5\times10^{6}$ nodes 
and $2.7\times 10^{7}$ links. 
\end{abstract}

% insert suggested PACS numbers in braces on next line

\maketitle
%\subsection{1. Introduction}

Many complex networks are organically evolving without any 
centralized control or design, and for this reason intense research has been 
devoted to understand their performance properties and more importantly, their 
vulnerabilities and failure modes.  In these studies, a fundamental role is played 
by centrality measures (originally introduced in social sciences  \cite{wasserman,
scott, sabidussi,friedkin,SocNet_BE06}), and in particular 
betweenness centrality \cite{freeman77,borgatti,anthonisse,Sameet}.  
Betweenness centrality (BC) of a node (edge) is defined as the fraction of all 
geodesics (shortest paths) 
passing through that node (edge). Since transport tends to minimize the cost/time 
of the route from source to destination, geodesics, and hence centrality measures 
and their distributions will strongly determine overall transport performance. 
Interestingly, geodesics are not only important for network flows but also for 
structural connectivity: removing nodes (edges) with high centrality one obtains 
a rapid increase in  diameter, and eventually the structural breakup of 
the  graph. Analysis of traffic, or information flow \cite{borgatti,
vespignani-traceroute,PRE-int-explor,Goh,Sameet,Danila,Guimera}, network vulnerability in face of attacks 
\cite{holme},  cascading failures \cite{motter,vesp09} or epidemics \cite{vespignani}, 
all involve betweenness calculations. 

Unfortunately, computation of betweenness  is very costly  \cite{brandes,newman,
SocNet_B08,Guimera,Noh-Rieger,Danila} and for large networks with millions to billions 
of nodes it is near impossible, hence approximation methods are needed. 
Existing approximations \cite{brandes-approx,geisberger-approx}, however, 
are sampling based, and ill controlled. 

Here we show that when geodesics are restricted to a maximum length $L$, the 
corresponding range-limited $L$-betweenness (introduced by Borgatti and 
Everett as bounded-distance betweenness \cite{SocNet_BE06}) for large graphs 
assumes a characteristic scaling form as function of $L$. This scaling can then be 
used to predict the betweenness distribution in the (usually unattainable)
diameter limit, and with good approximation, to predict the ranking of nodes/edges 
by betweenness. Additionally, the range-limited method generates $l$-betweenness 
values {\em for all} nodes and edges and {\em for all} $1 \leq l \leq L$, providing
systematic information on geodesics on all length-scales. This is of interest in its 
own right, when the transported entity has a small transmission probability (rumors, 
viruses) and thus high attrition rate, not exploring longer geodesics. As we show, 
the $L$-betweenness scaling is already  achieved for relatively small $L$ values 
and there is increasingly less new information obtained on BC distribution and 
ranking when going from $L$ to $L+1$.  The computational overhead, however,  involved 
in the $L \mapsto L+1$ step is usually immense. The range-limited centrality algorithm 
presented here, even in the diameter limit ($L = D$), has no larger complexity than 
the currently known fastest algorithms by Brandes \cite{brandes} and Newman 
\cite{newman}, that is $O(NM)$, where $N$ is the number of nodes 
and $M$ is the number of (directed) edges, and it is fully parallelizable. 
For $L < D$ our  algorithm runs {\em sublinearly} 
in $O(NM)$, making it possible to study networks with millions of  nodes. 
As an illustration, we analyzed a social network (SocNet) inferred from mobile 
phone trace-logs  \cite{Barabasi-mobility} having  a giant cluster with $N=5,568,785$  and 
$M=26,822,764$.  For this network we calculated all $L$-betweenness 
centralities ($L$-BCs) for all nodes and 
edges up to $L=5$ in $6$ days, on $10$ processors. With increasing $L$ the ranking 
of the highest BC nodes freezes and one can predict the top nodes early. The 
number of geodesics running through these nodes, however, explodes with  $L$. 
 For example, while the node with highest centrality for $L=4$ has $40,084,702$ geodesics, 
for $L=5$ it has $500,903,498$ of them passing through.

Calculating betweenness centrality of a node or edge in a directed graph 
$\bm{G}(V,\bm{E})$ requires to count the number of all-pair shortest directed 
paths incident on it. Here we include end-points, however, the algorithm can easily  
be changed to exclude them, or produce other variants.  The stress centrality (SC) 
$S(i)$ of a node $i \in V$ is simply the sum of the total number $\sigma_{mn}(i)$ of 
shortest directed paths from node $m$ to $n$ going through $i$,  $S(i)=
\sum_{m, n \in V}\sigma_{mn}(i)$.  Betweenness centrality (BC) \cite{freeman77,
anthonisse} normalizes the number of paths through a node by the total number 
of paths ($\sigma_{mn}$) for a given source-destination pair $(m,n)$:
$B(i)=\sum_{m, n \in V}\sigma_{mn}(i)/ \sigma_{mn}$. Similar quantities can be 
defined for an edge $(j,k) \in \bm{E}$: $S(j,k) = \sum_{m, n \in V}\sigma_{mn}(j,k)$ 
and $B(j,k)=\sum_{m, n \in V}\sigma_{mn}(j,k)/ \sigma_{mn}$.
\begin{figure}[htbp] \begin{center}
\includegraphics[width=0.36\textwidth]{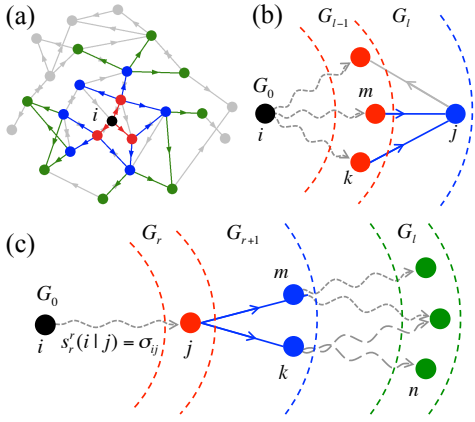}
\caption{a) Shells of the $\bm{C}_3$ subgraph of node $i$ (black) are colored 
red, blue, green. Grey elements are not part of the subgraph. b) Eq. (\ref{r1}) 
calculates SC of a node in $G_l$ (blue) by summing the SC of all its 
predecessors from $G_{l-1}(i)$ (red), e.g., $s_l^l(i|j)=s_{l-1}^{l-1}(i|k)+
s_{l-1}^{l-1}(i|m)$. c) Eqs.(\ref{r2}),(\ref{r3}) are based on the observations: 
$\sigma_{in}(j,k)=s_r^r(i|j) \sigma_{kn}$ and $\sigma_{in}(k)=s_{r+1}^{r+1}(i|k)
\sigma_{kn}$. Eq.(\ref{r4}) calculates the fixed-$l$ centralities for a node (red) 
in $G_r(i)$ by summing the corresponding centralities of its outgoing links 
(blue) in $G_{r+1}(i)$, e.g., $s_l^r(i|j)=s_l^{r+1}(i|j,k)+s_l^{r+1}(i|j,m)$.}\label{fig1} 
\vspace*{-0.5cm} \end{center} \end{figure}

In order to define range-limited quantities, let $s_l(j)$ and $b_l(j)$  denote the 
stress and betweenness centralities of a  node $j$ for all-pair shortest directed 
paths of {\em fixed} length $l$. Then $S_L(j) = \sum_{l=1}^L s_l(j)$ and 
$B_L(j) = \sum_{l=1}^L b_l(j)$ represent centralities from paths not longer than 
$L$. Similar measures for an edge are defined in the same way. Just as virtually 
all centrality algorithms, our method calculates these quantities for a node $j$ 
for shortest directed paths all emanating from a ``root'' node $i$, then it sums 
the obtained values for all $i \in V$  to get the final centralities for $j$ (similarly 
for edges).    
While the basic concept of our algorithm is similar to  Brandes' \cite{brandes} 
and Newman's \cite{newman}, we derive recursions that simultaneously compute 
both SC and BC for {\em all} nodes and edges and for all values $l=1,\ldots,L$. 
The algorithm's output thus generates detailed and systematic information 
about shortest paths in a graph on all length-scales, providing a tool for 
multiscale network analysis.

The algorithm starts from a given root $i$ and builds the $L$-range subgraph 
$\bm{C}_L$  containing all nodes which can be reached  in at most $L$ steps
from $i$. Only links which are part of the shortest paths starting from the root are 
included in $\bm{C}_L$. We decompose $\bm{C}_L$ into shells $G_l(i)$ containing  
all the nodes at shortest path distance $l$  from the root, and all incoming edges
from shell $l-1$, Fig.~\ref{fig1}a). The root itself is considered to be shell $0$ 
($G_0(i)$).

Let $s_{l}^{r}(i|j) = \sum_{n\in G_l}\sigma_{in}(j)$ denote the number of shortest 
directed paths of length $l$ from the root through node $j$ in the $r$-th shell 
$j \in G_r(i)$, and let $s_{l}^{r}(i|j,k) = \sum_{n\in G_l}\sigma_{in}(j,k)$ describe 
the same quantity for an edge $(j,k)$ in the $r$-th shell, $(j,k) \in G_r(i)$. We define 
similar quantities for betweenness, as $b_{l}^{r}(i|j) =  \sum_{n\in G_l}\sigma_{in}(j)/
\sigma_{in}$, and $b_{l}^{r}(i|j,k) = \sum_{n\in G_l}\sigma_{in}(j,k)/\sigma_{in}$.
Then $s_l(j) = \sum_{i\in V} s_{l}^{r}(i|j)$ and $b_l(j) = \sum_{i\in V} b_{l}^{r}(i|j)$,
with similar equations for edges. In these sums $r$ is not an independent variable.
Given $i$ and $j$, it is the radius of  shell $G_r(i)$ centered on $i$ and containing $j$.
One can show that the following recursions hold, (see also Fig.~\ref{fig1}):
\begin{eqnarray}
&&\!\!\!\!\!\! s_{l}^{l}(i|j) = \mbox{$\sum_{k}$} 
s_{l-1}^{l-1}(i|k)\;,\; b_{l}^{l}(i|j)=1, \label{r1}\\
&&\!\!\!\!\!\! s_{l}^{r+1}(i|j,k) = s_{l}^{r+1}(i|k) 
s_{r}^{r}(i|j)/s_{r+1}^{r+1}(i|k), \label{r2}\\
&&\!\!\!\!\!\! b_{l}^{r+1}(i|j,k) = b_{l}^{r+1}(i|k) 
s_{r}^{r}(i|j)/s_{r+1}^{r+1}(i|k), \label{r3}\\
&&\!\!\!\!\!\! s_{l}^{r}(i|j) = \mbox{$\sum_{k}$}  
s_{l}^{r+1}(i|j,k),\; b_{l}^{r}(i|j) = 
\mbox{$\sum_{k}$}  b_{l}^{r+1}(i|j,k).\label{r4}
\end{eqnarray}
The steps below are repeated for $l=1,\dots ,L$:
1) Build $G_l(i)$, using breadth-first search. 
2) Calculate the $l$-centrality measures ($s_{l}^{l}(i|j)$, $b_{l}^{l}(i|j)$) of all 
nodes in $G_l(i)$. 
3)  Moving backwards, through $r=l-1,...,1,0$, calculate the fixed-$l$ centralities 
of links in $G_{r+1}(i)$ and of nodes in $G_r(i)$, using recursions 
(\ref{r1}-\ref{r4}). Finally, return to step 1) until the last shell $G_L(i)$ is reached. 
In the end, we obtained the fixed-$l$ betweenness values of all nodes and edges in 
$\bm{C}_L$.  This concludes the basic algorithm, which can be modified 
to compute different variants of BC and SC, such as excluding endpoints. 
Similar recursions can also be derived for load and
closeness centrality \cite{SocNet_B08,borgatti}.

\begin{figure}[t] \begin{center}
\includegraphics[width=0.46\textwidth]{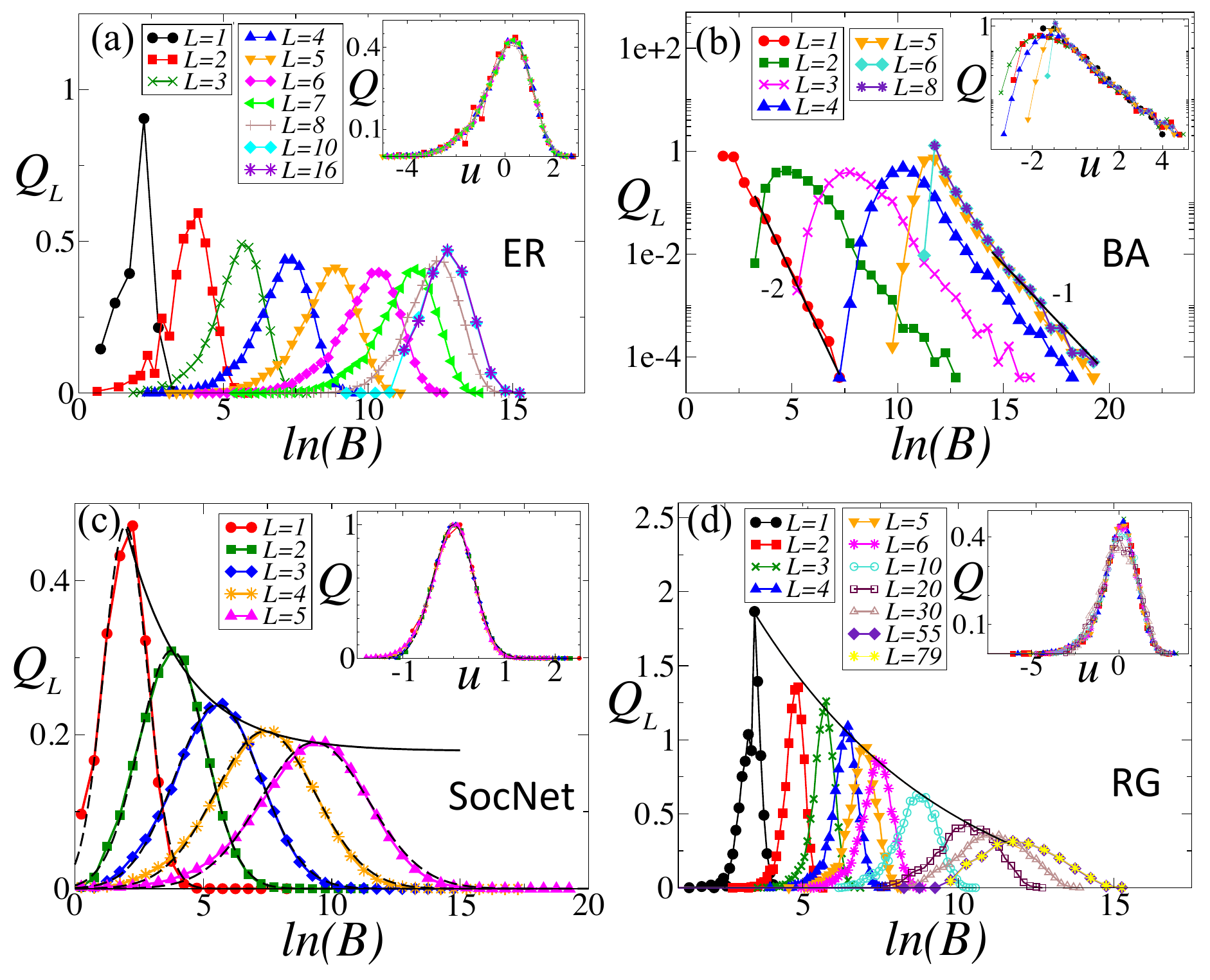}
\caption{Distribution $Q_L$ of $L$-betweeness for different values of  
$L$. a) ER, $N=5\times 10^4$, $\langle k\rangle=4$, diameter $D=16$, 
b) BA, $N=5\times 10^4$, $m=3$, $D=8$, 
c) SocNet, $N=5,568,785$, $M=26,822,764$ and the distributions are 
fitted by a lognormal (black dashed curves), d) RG, $N=10^4$, 
$\langle k\rangle =15$, $D=79$. The insets show the rescaled distributions, 
see text. } \label{distributions} \end{center} \vspace*{-0.75cm}\end{figure}

The $L$-betweenness values on large networks obey a scaling
behavior as function of $L$.  On Fig. \ref{distributions} we 
plot the distribution of node betweenness  values measured    on the 
Erd\H{o}s-R\'enyi (ER) random graph \cite{ER60}, the Barab\'asi-Albert (BA) 
scale-free model \cite{BA99}, the random geometric graph (RG) \cite{RG} and 
the large social network (SocNet) \cite{Barabasi-mobility}.
Since in large networks
$B_L$ grows quickly, it is better to work with the distribution 
$Q_L$ of the $\ln B_L$ values than with the distribution $P_L$ of $B_L$ values.
However, note that $Q_L(\ln B) = B P_L (B)$.
 As shown on the insets of Fig. \ref{distributions}, the distributions $Q_L(\ln B)$ for different 
$L$ can be rescaled onto each other by plotting $Q=\sigma_LQ_L $ vs
$u=[\ln(B)-\mu_L]/\sigma_L$, where $\mu_L$ and $\sigma_L$ are the mean 
and the standard deviation for  $\ln B_L$.  
These networks were chosen to represent very different graph classes: the ER, BA and 
SocNet have small diameters, while the RG has no shortcuts. The RG
is spatially embedded ($d=2$) unlike ER and BA; 
the SocNet, however, is influenced by the spatial 
embedding of people's motility \cite{Barabasi-mobility}. While BA has a power-law degree 
distribution $P(k)\sim k^{-3}$, both ER and RG have a Poissonian for
$P(k)$, and the SocNet's $P(k)$ resembles a log-normal 
\cite{lognormalSeshadri,Onnela}. Both RG and 
SocNet have high clustering, unlike the others.

Next we show that the scaling behavior observed for range-limited 
centralities in large graphs is a consequence of the scaling for shell sizes shown to exist
for e.g., in random graphs with arbitrary degree distributions \cite{newman_randomgraphs, shell_structure}.  
Here we present arguments for undirected, uncorrelated graphs and 
only deal with BC, extensions to other centralities mentioned above being 
straightforward. Let us define  $\langle\cdot\rangle$ as an average over all root 
nodes $i$ in the graph. If $z_{l}(i)$ denotes the number of nodes on 
shell $G_{l}(i)$, 
then we model the growth of shell sizes by a branching-like process 
$z_{l+1}(i)=z_{l}(i)\alpha_{l}\big[1+\epsilon_{l}(i)\big]$, where   
$\alpha_{l}=\langle z_{l+1}\rangle /\langle z_l \rangle$ is the
branching factor at an $l$-th shell, and $\epsilon_{l}(i)$ is a per-node, shell occupancy 
noise term, $|\epsilon_l|\ll1$, considered to obey $\langle \epsilon_{l}(i) \rangle = 0$ and 
$\langle \epsilon_{l}(i) \epsilon_{m}(i)\rangle = 2A_l \delta_{l,m}$, with 
$A_l$ decreasing with $l$, supported by numerical evidence. 
\begin{figure}[t] \begin{center}
\includegraphics[width=0.46\textwidth]{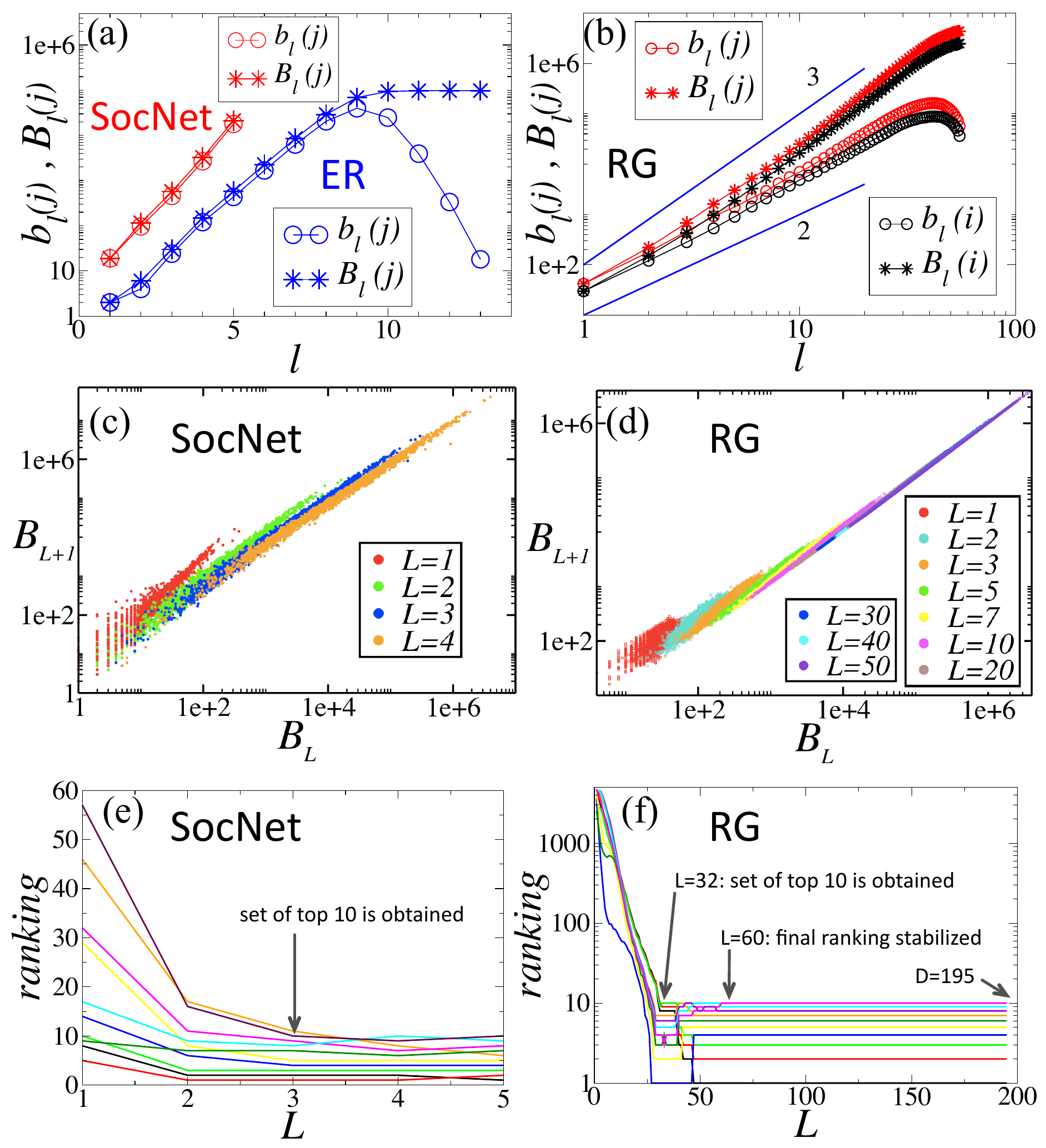}
\caption{  a) $b_l$ (circles)  and $B_l$ (stars) vs. $l$ for some
node $j$ in SocNet (red) and ER (blue). b) same as a) for RG for two arbitrary nodes $i$ and 
$j$. $B_{L+1}$ vs. $B_{L}$ for c) SocNet and d) 
RG. Each dot corresponds to a node. 
Ranking by BC vs $L$ for the top $10$ nodes in e) SocNet and f) 
RG (from Fig \ref{backbone}).}
\label{bBl}
\end{center} \vspace*{-0.5cm}\end{figure}
For undirected paths we can write $b_{l+1}(j)=(1/2)\sum_{i\in V}b_{l+1}(i|j) = 
z_{l+1}(j)+ (1/2)\sum_{m=1}^{l}\sum_{i\in G_m(j)} b_{l+1}^m(i|j) \equiv 
z_{l+1}(j)+ (1/2) u_{l+1}(j)$, where
we used the fact that in undirected graphs $i\in G_m(j) \Leftrightarrow j\in G_m(i)$.  
Note that the  number of terms in the inner sum $\sum_{i\in G_m(j)} b_{l+1}^m(i|j)$ 
is $z_m(j)$, which is rapidly increasing with $m$, and thus $u_{l+1}(j)$ is expected 
to have a weak dependence on $j$. Accordingly, we may approximate 
$ u_{l+1}(j) \simeq \sum_{m=1}^{l}\sum_{i\in G_m(j)} v^m_{l+1}(i)$, 
where $v^m_{l+1}(i)$ is an average betweenness computed on a shell of radius
$m$, {\em centered  on node} $i$ : $v^m_{l+1}(i)=\big[\sum_{k\in G_{m}(i)} 
b_{l+1}^m(i|k)\big]/z_{m}(i) $.
Based on the observation that $\sum_{k\in G_m(i)}b_l^m(i|k)=z_l(i)$, 
we can write that $v^m_{l+1}(i) \simeq z_{l+1}(i) /z_{m}(i)$. 
Using the recursion defined above for $z_{l+1}(i)$ as a branching process, 
and neglecting the small noise term, we obtain that $u_{l+1}(j) \simeq \alpha_l 
\sum_{m=1}^{l}  \sum_{i\in G_m(j)} z_{l}(i)/z_{m}(i)$. This allows  us to write a recursion 
 for $b_{l+1}(j)$ as  $b_{l+1}(j)\simeq \alpha_l [b_l(j)+ z_l(j)/2 +z_l(j)\epsilon_{l}(j)]$,
which can be iterated down to $l=1$, where $b_1(j)=z_1(j)=k_j$ is the degree of $j$:
\begin{equation}
b_{l}(j) \simeq \beta_{l} k_j e^{\xi_l(j)}\;, \label{nice}
\end{equation}
with 
$\beta_{l} = \frac{l+1}{2} \prod_{m=1}^{l-1}\alpha_m =  \frac{l+1}{2}
\langle z_{l}\rangle / \langle k \rangle$, and
$\xi_{l}(j)=\sum_{n=1}^{l-1}\!\frac{l+1-n}{l+1}\;\epsilon_{n}(j)$.
Eq (\ref{nice}) allows to relate the statistics of fixed-$l$ betweenness to the 
statistics of shell occupancies.  Since the noise term (calculated
from {\em per-node} occupancy deviations on a shell) is independent on root 
degree, the distribution of fixed-$l$ betweenness can be expressed as:
\begin{equation}
\rho_{l}(b) = 
\frac{1}{b}\int_1^{N-1}\!\!\!\!\!\!\!dk \;P(k) \Phi_l(\ln b - \ln \beta_l - \ln k)\;, \label{bd}
\end{equation}
where $P(k)$ is the degree distribution and $\Phi_l(\xi)$ is the distribution for the noise 
$\xi_l(j)$, peaked at $\xi = 0$, with fast decaying tails and $\Phi_1(x) = \delta(x)$. 
From (\ref{bd}) follows that 
the natural scaling variable for betweenness distribution is $u = \ln b - \ln \beta_l$. 
An extra $l$-dependence comes from the noise through the width $\sigma_l$ of 
$\Phi_l$ (for $l > 1$), which can be easily accounted for by the rescaling 
$u \mapsto u/\sigma_l$, $\rho_l \mapsto \rho_l \sigma_l$, collapsing the distributions 
for different $l$-values onto the same functional form. 
As $\Phi_l$ is sharply peaked around 0, the most significant contribution 
to the integral (\ref{bd}) for a given $b$ comes from degrees $k \simeq b/\beta_l$. 
Since $k \geq 1$, we have a rapid decay of $\rho_l(b)$ in the range $b < \beta_l$,
a maximum at $\overline{b} = \beta_l \overline{k}$ where $\overline{k}$ is the degree at
which $P(k)$ is maximum, and a sharp decay for $b > (N-1) \beta_l$.
In many networks, shell-size grows exponentially (ER, AB, and also in the SocNet), that is
$\alpha_l \simeq \alpha = \langle z_2 \rangle/\langle k \rangle$,  until  $l$
reaches the average shortest path distance. This  implies that $\beta_l\sim \alpha^l$ 
and $b_l$ grows exponentially with $l$ (Fig. \ref{bBl}a). 
In this case,  since $b_l$ is rapidly increasing with $l$, the cumulative 
$B_L(j)= \sum_{l} b_l(j)$ will be dominated by the largest $l$ values and thus, 
$B_L$ obeys a similar scaling supporting the observations in Fig. \ref{distributions}.
For pure scale-free networks $P(k) = ck^{-\gamma}$, and $\rho_l(b) \propto 
(b/\beta_l)^{1-\gamma}$ for $l > 1$.
In networks where the shell size grows as a power law (spatially embedded 
networks without shortcuts), such as RG, roadways, etc., $\beta_l\sim  l^{d}$, where $d$ 
is the embedding dimension, $b_l(j)\sim l^{d}$ and $B_L \sim l^{d+1}$ (Fig.\ref{bBl}.b). 

As the contributions of the noise terms $\epsilon_l(j)$ to $\xi_l(j)$ 
coming from larger shells are decreasing with 
increasing $l$ (their weight decreases as $(l+1)^{-1}$ in addition 
to the decreasing of their magnitude $|\epsilon_l(j)|$) the $\xi_l(j)$ quantities rapidly
converge to a constant.  From (\ref{nice}), for a pair of nodes $i,j$: $\ln [b_l(i)/b_l(j)]=\ln (k_i/k_j)+\xi_l(i)-\xi_l(j)$ showing that their relative ranking by $l$-betweenness freezes with increasing $l$. Consequently, $B_L$ and $B_{L+1}$ become more correlated with increasing $L$ 
(Fig\ref{bBl}c,d) and the ranking of the nodes by their BC also freezes (Fig.\ref{bBl}e,f), allowing
early prediction of top betweenness nodes.  Spatially embedded networks (RG) without shortcuts represent the worst case, but relative to their diameter the convergence of ranking is still fast (Fig.\ref{bBl}f). 
\begin{figure}[t] \begin{center}
\includegraphics[width=0.43\textwidth]{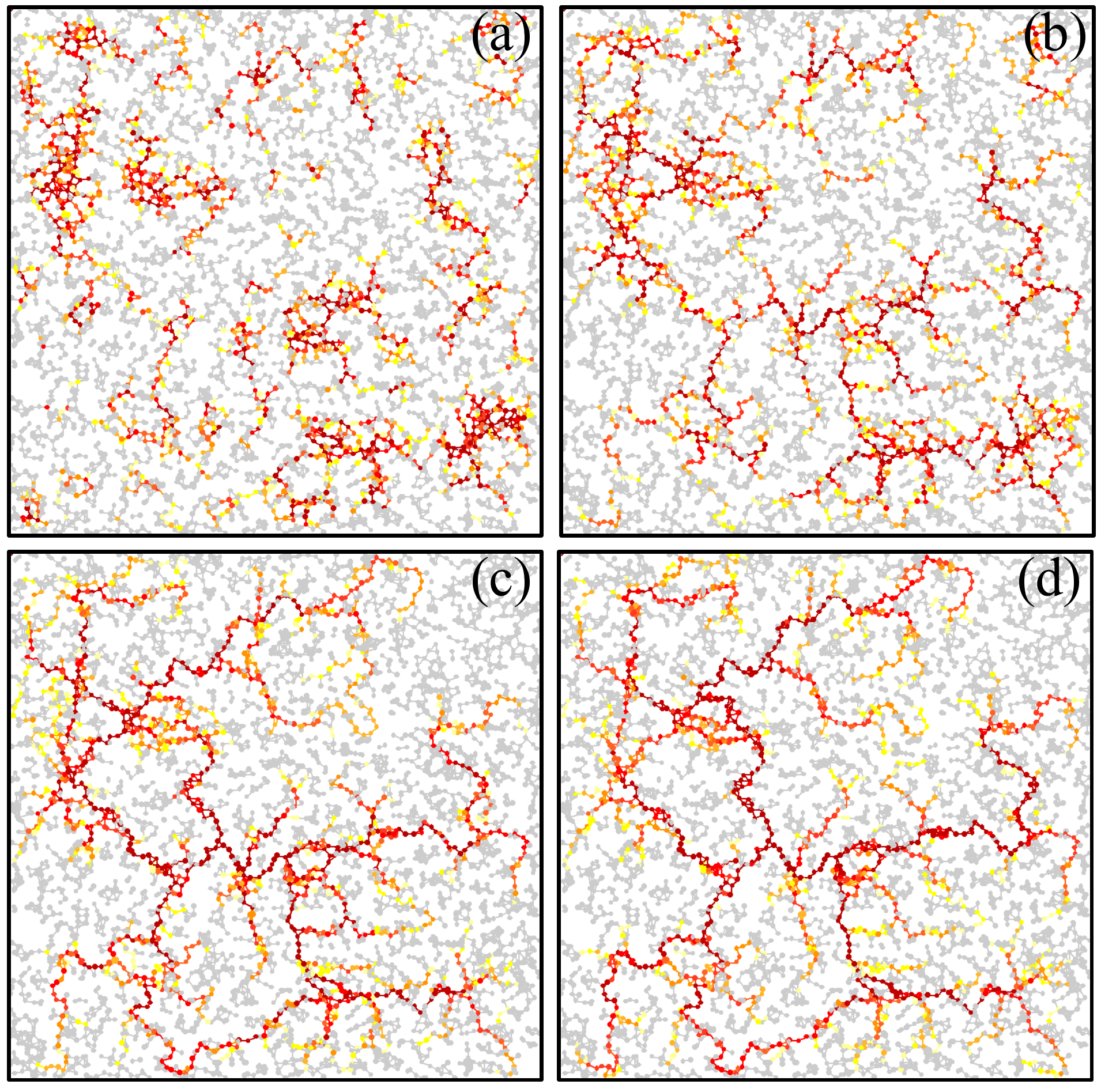}
\caption{Vulnerability backbone in a RG graph ($N=5\times10^3$, $\langle k\rangle=5$) for 
a) $L=5$, b) $L=15$, c) $L=45$, d) $L=D=195$. Darker red indicates nodes with higher $B_L$. 
In agreement with Fig.\ref{bBl}f,  VB is already well approximated at $L=45$, c).}
\label{backbone} \end{center} \vspace*{-0.65cm}\end{figure}
An important application of top betweenness predictability is  determining  the 
 ``vulnerability backbone'' (VB)  of a graph (crucial for network defense purposes \cite{holme,vespignani}) 
which is made by the smallest fraction of  highest betweenness nodes 
forming a percolating cluster through the network. Fig.\ref{backbone} for RG (worst case) 
shows that the VB (red subgraph) can accurately be predicted already from $L=45$ 
betweenness values (Fig.\ref{backbone}c) compared to the diameter ($D=195)$ based full 
betwennesses (Fig.\ref{backbone}d).   

Finally, we note that the scaling behavior can be used to provide a lower bound 
$L^*$ of the diameter, from
observing that finite size effects appear when the sum of average shell sizes 
hits $N$: $\sum_{l=1}^{L^*}\langle z_l \rangle =
\sum_{l=1}^{L^*}\frac{2}{l+1}\beta_l\langle k\rangle\simeq N$. This allows to find $L^*$
from the scaling behavior of $\beta_l$. In particular, for the  SocNet $L^* = 10$. 

In summary, we have shown that the contributions to centrality measures coming from
different length scales of the geodesics exhibit characteristic scaling in large graphs. 
Exploiting this universal property with the methods presented 
here makes it possible to predict betweenness values, distributions and ranking with 
relatively low computational costs.

\noindent This project was supported in part by the NSF BCS-0826958, 
HDTRA 201473-35045 and by the
Army Research Laboratory, W911NF-09-2-0053.  Views and conclusions are those of the authors, not representing those of the ARL or U.S. Govt.

%\bibliography{betw.bib}

\begin{thebibliography}{27}
\expandafter\ifx\csname natexlab\endcsname\relax\def\natexlab#1{#1}\fi
\expandafter\ifx\csname bibnamefont\endcsname\relax
  \def\bibnamefont#1{#1}\fi
\expandafter\ifx\csname bibfnamefont\endcsname\relax
  \def\bibfnamefont#1{#1}\fi
\expandafter\ifx\csname citenamefont\endcsname\relax
  \def\citenamefont#1{#1}\fi
\expandafter\ifx\csname url\endcsname\relax
  \def\url#1{\texttt{#1}}\fi
\expandafter\ifx\csname urlprefix\endcsname\relax\def\urlprefix{URL }\fi
\providecommand{\bibinfo}[2]{#2}
\providecommand{\eprint}[2][]{\url{#2}}

\bibitem[{\citenamefont{Wasserman and Faust}(1994)}]{wasserman}
\bibinfo{author}{\bibfnamefont{S.}~\bibnamefont{Wasserman}} \bibnamefont{and}
  \bibinfo{author}{\bibfnamefont{K.}~\bibnamefont{Faust}},
  \emph{\bibinfo{title}{{Social Network Analysis: methods and applications}}}
  (\bibinfo{publisher}{Cambridge Univ. Press}, \bibinfo{year}{1994}).

\bibitem[{\citenamefont{Scott}(1991)}]{scott}
\bibinfo{author}{\bibfnamefont{J.}~\bibnamefont{Scott}},
  \emph{\bibinfo{title}{{Social Network Analysis: A Handbook}}}
  (\bibinfo{publisher}{Sage Publications}, \bibinfo{year}{1991}).

\bibitem[{\citenamefont{Sabidussi}(1966)}]{sabidussi}
\bibinfo{author}{\bibfnamefont{G.}~\bibnamefont{Sabidussi}},
  \bibinfo{journal}{Psychometrika} \textbf{\bibinfo{volume}{31}},
  \bibinfo{pages}{581} (\bibinfo{year}{1966}).

\bibitem[{\citenamefont{Friedkin}(1991)}]{friedkin}
\bibinfo{author}{\bibfnamefont{N.~E.} \bibnamefont{Friedkin}},
  \bibinfo{journal}{Amer. J. of Soc.} \textbf{\bibinfo{volume}{96}},
  \bibinfo{pages}{1478} (\bibinfo{year}{1991}).

\bibitem[{\citenamefont{Borgatti and Everett}(2006)}]{SocNet_BE06}
\bibinfo{author}{\bibfnamefont{S. P. Borgatti}} \bibnamefont{and}
  \bibinfo{author}{\bibfnamefont{M. G.} \bibnamefont{Everett}},
  \bibinfo{journal}{Soc. Netw.} \textbf{\bibinfo{volume}{28}},
  \bibinfo{pages}{466} (\bibinfo{year}{2006}).

\bibitem[{\citenamefont{Freeman}(1977)}]{freeman77}
\bibinfo{author}{\bibfnamefont{L.~C.} \bibnamefont{Freeman}},
  \bibinfo{journal}{Sociometry} \textbf{\bibinfo{volume}{40}},
  \bibinfo{pages}{35} (\bibinfo{year}{1977}).

\bibitem[{\citenamefont{Borgatti}(2005)}]{borgatti}
\bibinfo{author}{\bibfnamefont{S.~P.} \bibnamefont{Borgatti}},
  \bibinfo{journal}{Soc. Netw.} \textbf{\bibinfo{volume}{27}},
  \bibinfo{pages}{55} (\bibinfo{year}{2005}).

\bibitem[{\citenamefont{Anthonisse}(1971)}]{anthonisse}
\bibinfo{author}{\bibfnamefont{J.~M.} \bibnamefont{Anthonisse}},
  \bibinfo{journal}{Tech. Rep. BN 9/71, Stichting Math. Centr., Amsterdam}
  (\bibinfo{year}{1971}).

\bibitem[{\citenamefont{Sreenivasan et~al.}(2002)\citenamefont{Sreenivasan,Cohen, Lopez,Toroczkai}}]{Sameet}
\bibinfo{author}{\bibfnamefont{S.}~\bibnamefont{Sreenivasan \textit{et al.}}},
  \bibinfo{journal}{Phys. Rev. E} \textbf{\bibinfo{volume}{75}}, \bibinfo{pages}{036105}
  (\bibinfo{year}{2007}).


\bibitem[{\citenamefont{Dall\char39{}Asta
  et~al.}(2006)\citenamefont{Dall\char39{}Asta, Alvarez-Hamelin, Barrat,
  V\'azquez, and Vespignani}}]{vespignani-traceroute}
\bibinfo{author}{\bibfnamefont{L.}~\bibnamefont{Dall\char39{}Asta \textit{et al.}}},
  \bibinfo{journal}{Theor. Comp. Sci.} \textbf{\bibinfo{volume}{355}},
  \bibinfo{pages}{6} (\bibinfo{year}{2006}).

\bibitem[{\citenamefont{Dall\char39{}Asta
  et~al.}(2005)\citenamefont{Dall\char39{}Asta, Alvarez-Hamelin, Barrat,
  V\'azquez, and Vespignani}}]{PRE-int-explor}
\bibinfo{author}{\bibfnamefont{L.}~\bibnamefont{Dall\char39{}Asta \textit{et al.}}},
  \bibinfo{journal}{Phys. Rev. E} \textbf{\bibinfo{volume}{71}},
  \bibinfo{pages}{036135} (\bibinfo{year}{2005}).


\bibitem[{\citenamefont{Goh et~al.}(2002)\citenamefont{Goh,Kahng, Kim}}]{Goh}
\bibinfo{author}{\bibfnamefont{K.-I.}~\bibnamefont{Goh \textit{et al.}}},
  \bibinfo{journal}{Phys. Rev. Lett.} \textbf{\bibinfo{volume}{87}}, \bibinfo{pages}{278701}
  (\bibinfo{year}{2001}).


\bibitem[{\citenamefont{Danila et~al.}(2002)\citenamefont{Danilla,Yu, Earl,Toroczkai}}]{Danila}
\bibinfo{author}{\bibfnamefont{B.}~\bibnamefont{Danila \textit{et al.}}},
  \bibinfo{journal}{Phys. Rev. E} \textbf{\bibinfo{volume}{74}}, \bibinfo{pages}{046114}
  (\bibinfo{year}{2006}).


\bibitem[{\citenamefont{Guimera et~al.}(2002)\citenamefont{Guimera,Diaz-Guilera, Vega-Redondo,Cabrales,Arenas}}]{Guimera}
\bibinfo{author}{\bibfnamefont{R.}~\bibnamefont{Guimer\`a \textit{et al.}}},
  \bibinfo{journal}{Phys. Rev. Lett.} \textbf{\bibinfo{volume}{89}}, \bibinfo{pages}{248701}
  (\bibinfo{year}{2001}).


\bibitem[{\citenamefont{Holme et~al.}(2002)\citenamefont{Holme, Kim, Yoon, and
  Han}}]{holme}
\bibinfo{author}{\bibfnamefont{P.}~\bibnamefont{Holme \textit{et al.}}},
  \bibinfo{journal}{Phys. Rev. E} \textbf{\bibinfo{volume}{65}}, \bibinfo{pages}{056109}
  (\bibinfo{year}{2002}).

\bibitem[{\citenamefont{Motter}(2002)\citenamefont{Motter}}]{motter}
\bibinfo{author}{\bibfnamefont{A.E.}~\bibnamefont{Motter}},
  \bibinfo{journal}{Phys. Rev. Lett.} \textbf{\bibinfo{volume}{93}},
  \bibinfo{pages}{098701}
  (\bibinfo{year}{2004}).
  
  \bibitem[{\citenamefont{Vespignani}(2002)\citenamefont{Vespignani}}]{vesp09}
\bibinfo{author}{\bibfnamefont{A.}~\bibnamefont{Vespignani}},
  \bibinfo{journal}{Science} \textbf{\bibinfo{volume}{325}},
  \bibinfo{pages}{425}
  (\bibinfo{year}{2009}).

\bibitem[{\citenamefont{DallAsta et~al.}(2006)\citenamefont{DallAsta, Barrat,
  Barthelemy, and Vespignani}}]{vespignani}
\bibinfo{author}{\bibfnamefont{L.}~\bibnamefont{Dall'Asta \textit{et al.}}},
    \bibinfo{journal}{J.Stat.Mech.},
    \bibinfo{pages}{P04006},
  (\bibinfo{year}{2006}).

\bibitem[{\citenamefont{Brandes}(2001)}]{brandes}
\bibinfo{author}{\bibfnamefont{U.}~\bibnamefont{Brandes}}, \bibinfo{journal}{J.
  of Math. Sociology} \textbf{\bibinfo{volume}{25}}, \bibinfo{pages}{163}
  (\bibinfo{year}{2001}).

\bibitem[{\citenamefont{Newman}(2001)}]{newman}
\bibinfo{author}{\bibfnamefont{M.~E.~J.} \bibnamefont{Newman}},
  \bibinfo{journal}{Phys. Rev. E} \textbf{\bibinfo{volume}{64}},
  \bibinfo{pages}{016132} (\bibinfo{year}{2001}).

\bibitem[{\citenamefont{Brandes}(2008)}]{SocNet_B08}
\bibinfo{author}{\bibfnamefont{U.}~\bibnamefont{Brandes}},
  \bibinfo{journal}{Soc. Netw.} \textbf{\bibinfo{volume}{30}},
  \bibinfo{pages}{136} (\bibinfo{year}{2008}).


\bibitem[{\citenamefont{Noh,Rieger}(2008)}]{Noh-Rieger}
\bibinfo{author}{\bibfnamefont{J.D.}~\bibnamefont{Noh} \bibnamefont{and} \bibnamefont{H.}~\bibnamefont{Rieger}},
  \bibinfo{journal}{Phys. Rev. Lett.} \textbf{\bibinfo{volume}{92}},
  \bibinfo{pages}{118701} (\bibinfo{year}{2004}).

\bibitem[{\citenamefont{Brandes and Pich}(2007)}]{brandes-approx}
\bibinfo{author}{\bibfnamefont{U.}~\bibnamefont{Brandes}} \bibnamefont{and}
  \bibinfo{author}{\bibfnamefont{C.}~\bibnamefont{Pich}}, \bibinfo{journal}{I.
  J. Bif. Chaos} \textbf{\bibinfo{volume}{17}}, \bibinfo{pages}{2303}
  (\bibinfo{year}{2007}).

\bibitem[{\citenamefont{Geisberger et~al.}(2008)\citenamefont{Geisberger,
  Sanders, and Schultes}}]{geisberger-approx}
\bibinfo{author}{\bibfnamefont{R.}~\bibnamefont{Geisberger \textit{et al.}}}, in
  \emph{\bibinfo{booktitle}{ALENEX}} , 
  \bibinfo{pages}{90} (\bibinfo{year}{2008}).

\bibitem[{\citenamefont{Gonz\'alez et~al.}(2008)\citenamefont{Gonz\'alez,
  Hidalgo, and Barab\'asi}}]{Barabasi-mobility}
\bibinfo{author}{\bibfnamefont{M.~C.} \bibnamefont{Gonz\'alez \textit{et al.}}},
  \bibinfo{journal}{Nature}
  \textbf{\bibinfo{volume}{453}}, \bibinfo{pages}{779} (\bibinfo{year}{2008}).

\bibitem[{\citenamefont{Erd\H{o}s and R\'enyi}(1960)}]{ER60}
\bibinfo{author}{\bibfnamefont{P.}~\bibnamefont{Erd\H{o}s}} \bibnamefont{and}
  \bibinfo{author}{\bibfnamefont{A.}~\bibnamefont{R\'enyi}},
  \bibinfo{journal}{Publ. Math. Inst. Hung. Acad. Sci}
  \textbf{\bibinfo{volume}{5}}, \bibinfo{pages}{17} (\bibinfo{year}{1960}).

\bibitem[{\citenamefont{Barab\'asi and Albert}(1999)}]{BA99}
\bibinfo{author}{\bibfnamefont{A.~L.} \bibnamefont{Barab\'asi}}
  \bibnamefont{and} \bibinfo{author}{\bibfnamefont{R.}~\bibnamefont{Albert}},
  \bibinfo{journal}{Science} \textbf{\bibinfo{volume}{286}},
  \bibinfo{pages}{509} (\bibinfo{year}{1999}).

\bibitem[{\citenamefont{Dall and Christensen}(2002)}]{RG}
\bibinfo{author}{\bibfnamefont{J.}~\bibnamefont{Dall}} \bibnamefont{and}
  \bibinfo{author}{\bibfnamefont{M.}~\bibnamefont{Christensen}},
  \bibinfo{journal}{Phys. Rev. E} \textbf{\bibinfo{volume}{66}},
  \bibinfo{pages}{016121} (\bibinfo{year}{2002}).

\bibitem[{\citenamefont{Newman et~al.}(2001)\citenamefont{Newman, Strogatz, and
  Watts}}]{newman_randomgraphs}
\bibinfo{author}{\bibfnamefont{M.~E.~J.}~\bibnamefont{Newman \textit{et~al.}}},
  \bibinfo{journal}{Phys. Rev. E} \textbf{\bibinfo{volume}{64}},
  \bibinfo{pages}{026118} (\bibinfo{year}{2001}).

\bibitem[{\citenamefont{Shao et~al.}(2009)\citenamefont{Shao, Buldyrev,
  Braunstein, Havlin, and Stanley}}]{shell_structure}
\bibinfo{author}{\bibfnamefont{J.}~\bibnamefont{Shao \textit{et al.}}},
  \bibinfo{journal}{Phys. Rev. E} \textbf{\bibinfo{volume}{80}},
  \bibinfo{pages}{036105} (\bibinfo{year}{2009}).

   \bibitem[{\citenamefont{Onnela et~al.}(2007)\citenamefont{Onnela et~al.}}]{Onnela}
\bibinfo{author}{\bibfnamefont{J. P.}~\bibnamefont{Onnela \textit{et al.}}},
  \bibinfo{journal}{PNAS}, \textbf{\bibinfo{volume}{104}},
  \bibinfo{pages}{7332} (\bibinfo{year}{2007}).

 \bibitem[{\citenamefont{Seshadri et~al.}(2008)\citenamefont{Seshadri, Machiraju,
  Sridharan, Bolot, Faloutsos and Leskove}}]{lognormalSeshadri}
\bibinfo{author}{\bibfnamefont{M.}~\bibnamefont{Seshadri \textit{et al.}}},
  \bibinfo{journal}{SIGKDD-08}
   (\bibinfo{year}{2008}).


\end{thebibliography}

\end{document}